\documentclass[runningheads]{llncs}

\usepackage[utf8]{inputenc}
\usepackage[T1]{fontenc}
\usepackage{graphicx}
\usepackage{amsmath,amssymb,amsfonts}
\usepackage{url}
\usepackage{booktabs}
\usepackage{tabularx}
\usepackage{array}
\usepackage{caption}
\usepackage{rotating}
\usepackage{multirow}
\usepackage{tikz}
\usetikzlibrary{arrows.meta, positioning, fit}
\usepackage{siunitx}
\usepackage[numbers,sort&compress]{natbib}

\DeclareSIUnit\dBm{dBm}
\DeclareSIUnit\byte{B}
\DeclareSIUnit\hertz{Hz}

\begin{document}

\title{Resilience Analysis in Off-Grid LoRa Mesh Networks: Evaluation of
Meshtastic Profiles in Long-Range Propagation Scenarios}

\author{Guillermo Antonio Hernandez Ortiz\inst{1} \and
Edgar Santiago Quiroz Puentes\inst{1} \and
Jose de Jes\'us Rugeles\inst{1}}

\authorrunning{G.\,A. Hernandez Ortiz et al.}

\institute{Telecommunications Engineering Program, Universidad Militar Nueva
Granada, Bogot\'a D.C., Colombia\\
\email{est.guillermo.hern@unimilitar.edu.co,
est.edgar.squiroz@unimilitar.edu.co,
jose.rugeles@unimilitar.edu.co}}

\maketitle

\begin{abstract}
The integration of LoRa technologies with mesh topologies represents a robust
alternative for off-grid communications in emergency scenarios within smart
cities. Meshtastic firmware implements a decentralised mesh network over LoRa
where each node acts simultaneously as end device and router, enabling
communication via Bluetooth-connected mobile devices without reliance on
conventional infrastructure. Within the Colombian context
(\SI{915}{\mega\hertz} ISM band), this work establishes design and planning
criteria through a controlled guided-link methodology that isolates the LoRa
physical layer from propagation effects, enabling deterministic
characterisation of all eight Meshtastic modem presets at three transmission
power levels (42 datasets). The results reveal a performance partitioning
governed primarily by Spreading Factor~(SF): \textit{Short} presets
(SF7--SF8) fail at \SIrange{110}{120}{\decibel} of path attenuation,
\textit{Medium} presets (SF9--SF10) sustain links up to
\SIrange{135}{150}{\decibel}, and \textit{Long} presets (SF11--SF12) maximise
coverage, with \textit{Long Slow} reaching \SI{180}{\decibel} before
failure---a \SIrange{60}{70}{\decibel} advantage over the fastest profiles.
The SNR analysis confirms sub-noise-floor demodulation down to
\SI{-18}{\decibel} for SF12, with abrupt link failure occurring within
\SIrange{2}{4}{\decibel} of the theoretical limit. Based on these thresholds,
three operational regimes are defined---high-density IoT, balanced urban mesh,
and maximum-range emergency---providing network designers with quantitative
criteria to select the appropriate configuration and node density for smart
city deployments.

\end{abstract}

\section{Introduction}

The transformation of urban environments into smart cities fundamentally
depends on the ability to deploy communication networks that connect thousands
of devices with limited energy, memory, and processing resources. In this
ecosystem, Low Power Wide Area Networks (LPWAN) based on LoRa (Long Range)
modulation have emerged as a technically compelling solution, filling the gap
between short-range networks such as WiFi and traditional cellular
infrastructure. Through its Chirp Spread Spectrum (CSS) modulation, LoRa
provides superior resistance to interference and multipath fading, which is
critical for data transmission in dense metropolitan
environments~\cite{augustin_study_2016}.

Traditionally, these solutions have been deployed under the LoRaWAN protocol
using a star topology, where end devices communicate with centralised gateways
connected to Internet backhaul. However, disaster events---earthquakes, floods,
massive power outages, or cyberattacks on critical infrastructure---can disable
the 4G/5G cellular and fibre optic networks that support urban services such as
transportation, energy distribution, and public
safety~\cite{paikaray_comprehensive_2024}. This reality has driven the
development of decentralised mesh network architectures capable of maintaining
communication continuity without any fixed infrastructure, a factor identified
as a fundamental pillar of urban resilience in the context of disaster-resilient
smart city frameworks~\cite{lai_review_2020, yuloskov_smart_2021}.

The Meshtastic firmware implements precisely this approach over LoRa, enabling
every node to function simultaneously as an end device and a router through a
managed flooding protocol combined with CSMA/CA. By enabling Bluetooth Low
Energy (BLE) connectivity, users can communicate through Meshtastic nodes using
their mobile devices without depending on conventional cellular or Internet
infrastructure. Beyond emergency messaging, the protocol natively supports
telemetry data transmission from IoT sensors, while compatibility with MQTT
allows Internet-connected nodes to act as gateways to urban management
platforms such as ThingsBoard or Grafana~\cite{zhou_design_2019}. This hybrid
architecture enables autonomous operation during infrastructure failures,
bridging the gap between emergency communications and smart city services such
as waste management, smart metering, and real-time structural health monitoring
of bridges and critical infrastructure.

In the context of Latin American smart cities, where telecommunications
infrastructure can be heterogeneous and vulnerable, the ability to operate in
the \SI{915}{\mega\hertz} ISM band without requiring a licence, combined with
low-cost hardware and solar-powered autonomous operation, makes Meshtastic a
particularly compelling solution for community early warning systems. The
present work seeks to establish a design and action plan for the use of
Meshtastic mesh networks in the \SI{915}{\mega\hertz} frequency band, with
the goal of contributing to the development of resilient, low-cost
communication infrastructure for smart urban environments.

The objective of this work is to establish design and planning criteria for
Meshtastic networks operating in the \SI{915}{\mega\hertz} ISM band. To this
end, this paper makes the following contributions:
\begin{enumerate}
    \item A controlled guided-link experimental methodology that isolates LoRa
    physical layer behaviour from propagation effects, enabling deterministic
    sensitivity and PER characterisation.

    \item A comprehensive evaluation of all eight Meshtastic modem presets
    (\textit{Short Turbo} through \textit{Long Slow}) at three transmission
    power levels, yielding 42 datasets that map the operational envelope of
    the SX1262 transceiver.

    \item Spectral verification of LoRa emissions using both SDR-based and
    calibrated spectrum analyser measurements, confirming CSS modulation
    compliance in the \SI{915}{\mega\hertz} band.

    \item Practical design guidelines for coverage planning in urban emergency
    and IoT deployment scenarios, derived from empirical sensitivity
    thresholds.
\end{enumerate}

The remainder of this paper is organised as follows.
Section~\ref{sec:methodology} details the experimental setup, hardware
characterisation, and test matrix. Section~\ref{sec:results} presents the
measurement results and sensitivity analysis. Section~\ref{sec:discussion}
discusses the implications for configuration selection and node density
planning. Finally, Section~\ref{sec:conclusions} summarises the findings and
outlines future work.

\section{State of the art}
\label{sec:State of the art }

The development of smart cities requires communication technologies that achieve a balance between long range and low energy consumption, especially when connecting devices with limited resources. In this context, LoRa (Long Range) has established itself as one of the most relevant solutions at the physical layer, thanks to its Chirp Spread Spectrum (CSS) modulation, which allows it to operate robustly against interference and fading effects typical of dense urban environments. This characteristic translates into high receiver sensitivity, making it possible to establish links where other wireless technologies are not viable.
A key aspect of these networks is energy efficiency. Mechanisms such as Channel Activity Detection (CAD) allow devices to identify channel activity without keeping the receiver continuously active, thereby reducing battery consumption. However, parameters such as the Spreading Factor (SF) and bandwidth (BW) not only determine communication range, but also the time on air (airtime), which directly impacts node battery life.

In relation to this, studies such as that of ~\cite{augustin_study_2016} analyze in detail the behavior of the LoRa physical layer, establishing the mathematical foundations for understanding how SF and BW influence metrics such as bit rate and symbol duration. Their results show that SF is the most determining factor in coverage, which is consistent with experimental observations in practical configurations such as those used in Meshtastic.

On the other hand, ~\cite{kurji_lora_2021} evaluate LoRa performance in environments with adverse propagation conditions, such as university campuses with multiple obstacles. Their results show how SF and Code Rate (CR) affect packet loss rate, providing a relevant experimental basis for analysis in dense urban scenarios. Along the same lines, ~\cite{zhou_design_2019} present real-world implementations of private LoRa networks, reporting ranges of up to 7.5~km in urban environments using high SF configurations, which serves as a reference for long-range propagation studies.

At the architecture level, most current deployments are based on LoRaWAN, which uses a star topology dependent on gateways and central servers. Nevertheless, alternatives such as mesh networks have emerged, enabling direct communication between nodes without the need for fixed infrastructure. This approach, adopted by solutions such as Meshtastic, is particularly attractive in scenarios where traditional connectivity is unavailable or may fail, such as in emergency situations.

The behavior of these networks under real conditions has been widely studied. ~\cite{blenn_lorawan_2017} analyzed the operation of The Things Network over several months, identifying issues associated with packet collisions and SF usage distribution in congested networks. Likewise,
~\cite{setiowati_lora_2023} demonstrate the applicability of LoRa in critical infrastructure monitoring, evaluating quality of service parameters such as latency and throughput, which reinforces its usefulness in urban environments.

From a broader perspective, systematic reviews such as that of ~\cite{paikaray_comprehensive_2024} show the growing adoption of LoRa in sectors such as agriculture, mining, and early warning systems. At the same time, ~\cite{lavric_internet_nodate} highlight important challenges related to security and interconnectivity in LoRaWAN architectures, which opens the door to decentralized solutions such as mesh networks. Regarding the regulatory framework, ~\cite{lai_review_2020} emphasize the importance of international standards such as ISO~37120 and IEEE~P2784, which promote interoperability and scalability in smart city development.

Finally, the deployment of these technologies is often supported by controlled testing environments. ~\cite{omotayo_systems_2021} introduce the concept of Living Labs, which allows solutions to be validated before large-scale implementation. In a complementary manner, studies such as those of ~\cite{yuloskov_smart_2021} and \cite{cantuarias-villessuzanne_clustering_2021} emphasize the importance of integrating sensors and resilient networks to improve key aspects such as urban mobility and public safety.

Taken together, the reviewed literature shows that the combination of LoRa with mesh architectures represents a solid alternative for the development of resilient networks, especially in scenarios where centralized infrastructure is not viable. This approach is particularly relevant in emergency communication applications and off-grid environments, where service continuity is a critical requirement.


\section{Methodology}
\label{sec:methodology}

The design of wireless networks using Meshtastic nodes requires a rigorous
understanding of radio frequency (RF) parameters and hardware capabilities.
This section details the experimental framework, the characterisation of the
devices under test (DUT), and the physical layer principles governing system
performance.

\subsection{Experimental Setup and System Architecture}

To evaluate link resilience and packet error rate (PER) deterministically, the
RF link was replaced by a guided medium. This approach eliminates stochastic
variables such as multipath fading, environmental interference, and temporal
variability of the wireless channel, providing a controlled and repeatable
environment for sensitivity analysis.

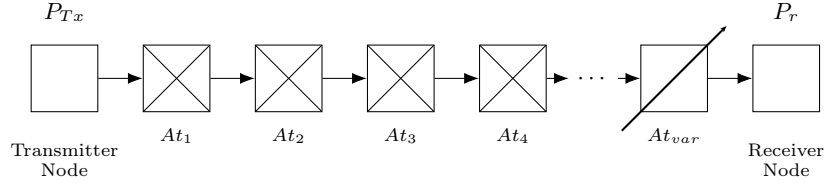
\begin{figure}[htbp]
  \centering
  \resizebox{0.9\textwidth}{!}{%
  \begin{tikzpicture}[
    node distance = 4mm and 6mm,
    block/.style      = {draw, minimum width=9mm, minimum height=9mm},
    att_fixed/.style  = {draw, minimum width=9mm, minimum height=9mm,
                         path picture={
                           \draw (path picture bounding box.north west)
                                 -- (path picture bounding box.south east);
                           \draw (path picture bounding box.north east)
                                 -- (path picture bounding box.south west);
                         }},
    att_var/.style    = {draw, minimum width=9mm, minimum height=9mm,
                         append after command={
                           \pgfextra{
                             \draw[-{Stealth[length=2.5pt,width=2pt]}, thick]
                               ([xshift=-2.5mm, yshift=-2.5mm]\tikzlastnode.south west)
                               -- ([xshift=2.5mm, yshift=2.5mm]\tikzlastnode.north east);
                           }
                         }},
    arr/.style = {-{Latex[length=2mm]}, thin},
  ]
    \node[block]                        (tx)    {};
    \node[att_fixed, right=of tx]       (at1)   {};
    \node[att_fixed, right=of at1]      (at2)   {};
    \node[att_fixed, right=of at2]      (at3)   {};
    \node[att_fixed, right=of at3]      (at4)   {};
    \node[right=3mm of at4, font=\small](dots)  {$\cdots$};
    \node[att_var,   right=3mm of dots] (atvar) {};
    \node[block,     right=of atvar]    (rx)    {};

    \foreach \a/\b in {tx/at1, at1/at2, at2/at3, at3/at4, dots/atvar, atvar/rx}
      \draw[arr] (\a) -- (\b);
    \draw[arr] (at4.east) -- (dots.west);

    \node[above=2mm of tx,    font=\small]      {$P_{Tx}$};
    \node[above=2mm of rx,    font=\small]      {$P_r$};
    \node[below=1mm of at1,   font=\scriptsize] {$At_1$};
    \node[below=1mm of at2,   font=\scriptsize] {$At_2$};
    \node[below=1mm of at3,   font=\scriptsize] {$At_3$};
    \node[below=1mm of at4,   font=\scriptsize] {$At_4$};
    \node[below=1mm of atvar, font=\scriptsize] {$At_{var}$};
    \node[below=3mm of tx, align=center, font=\scriptsize] {Transmitter\\Node};
    \node[below=3mm of rx, align=center, font=\scriptsize] {Receiver\\Node};
  \end{tikzpicture}%
  }
  \caption{Experimental guided-link system model using fixed attenuators
  ($At_{1\text{--}4}$: Mini-Circuits VAT-30+, \SI{50}{\ohm}) and a manual
  rotary step attenuator ($At_{var}$: 0--\SI{110}{\decibel}).}
  \label{fig:modelo-cascada-lora-final}
\end{figure}

Figure~\ref{fig:modelo-cascada-lora-final} illustrates the experimental link
model. The antennas were substituted by a wired connection where the RF channel
is emulated through a cascade of fixed attenuators ($At_{1\text{--}4}$) and a
final manual rotary step attenuator ($At_{var}$) used to sweep the receiver
toward its sensitivity threshold. All components are matched to a
\SI{50}{\ohm} characteristic impedance.

The precision attenuators used were Mini-Circuits model VAT-30+, which provide
a flat response within the \SI{915}{\mega\hertz} band with a
$VSWR \approx 1.2$. Interconnections were made using \SI{50}{\ohm}
SMA-to-SMA coaxial jumpers with measured insertion loss
$L_{ins} < \SI{0.5}{\decibel}$ per segment. The combined uncertainty of the
attenuator chain, including connector mismatch and component tolerances, is
estimated at $\pm\SI{0.8}{\decibel}$ across the measurement bandwidth.

The received power $P_r$ is calculated as:
\begin{equation}
    P_r = P_{Tx} - L_{ins} - \sum_{x=1}^{n} At_x
    \label{eq:potencia-recibida-cable}
\end{equation}
where $P_{Tx}$ is the transmission power, $L_{ins}$ represents the aggregate
insertion losses of connectors and jumpers, and $At_x$ denotes the attenuation
of each stage.

The study utilised the Heltec WiFi LoRa~32 V3.2 and the RAK4631 WisBlock as
primary nodes. Both utilise the Semtech SX1262 LoRa chipset. It is important
to note a discrepancy between software reporting and hardware limits: while the
Meshtastic interface for the RAK4631 node used as a transmitter may allow
settings up to $+\SI{30}{\dBm}$, the SX1262 transceiver is physically limited
to a maximum output of $+21 \pm \SI{1}{\dBm}$~\cite{semtech_corporation_sx1276777879_2020}.
Consequently, this study adopts qualitative labels (Low, Medium, Max) to
describe transmission power levels, where ``Max'' represents the hardware's
maximum effective saturation point. Table~\ref{tab:comparison-heltec-rak-expanded}
compares the key technical specifications of both nodes.

\begin{table}[htbp]
\centering
\caption{Technical Comparison of Meshtastic Nodes under Evaluation}
\label{tab:comparison-heltec-rak-expanded}
\small
\begin{tabular}{p{0.25\textwidth}p{0.32\textwidth}p{0.32\textwidth}}
\toprule
\textbf{Feature} & \textbf{Heltec WiFi LoRa~32 V3.2} & \textbf{RAK4631 WisBlock} \\
\midrule
Master MCU
  & ESP32-S3FN8 (Dual-core LX7, \SI{240}{\mega\hertz})
  & Nordic nRF52840 (Cortex-M4F, \SI{64}{\mega\hertz}) \\
LoRa Chipset   & Semtech SX1262 & Semtech SX1262 \\
LoRa Sensitivity
  & \SI{-137}{\dBm} (SF12 / \SI{125}{\kilo\hertz})
  & \SI{-137}{\dBm} (SF12 / \SI{125}{\kilo\hertz}) \\
Max.\ TX Power & $21 \pm \SI{1}{\dBm}$ & \SI{22}{\dBm} \\
Flash / RAM
  & \SI{8}{\mega\byte} Flash / \SI{512}{\kilo\byte} SRAM
  & \SI{1}{\mega\byte} Flash / \SI{256}{\kilo\byte} RAM \\
Onboard Display & 0.96'' OLED ($128 \times 64$) & None (external via I\textsuperscript{2}C) \\
USB Interface   & Type-C (CP2102 bridge) & Type-C (Native USB) \\
\bottomrule
\end{tabular}
\end{table}

\subsection{LoRa Physical Layer and Meshtastic Presets}

\subsubsection{Modem Presets (SF, BW, CR).}
Meshtastic abstracts the complexity of LoRa modulation through ``Modem
Presets,'' which are standardised combinations of three core variables:
\begin{itemize}
    \item \textbf{Spreading Factor (SF):} Defines the duration of the chirps.
    A higher SF (up to 12) increases the link budget and sensitivity, allowing
    signals to be decoded even under the noise floor, but significantly
    increases Time-on-Air (ToA).

    \item \textbf{Bandwidth (BW):} The frequency range of the signal. Reducing
    BW (e.g., from \SI{500}{\kilo\hertz} to \SI{125}{\kilo\hertz}) improves
    sensitivity at the cost of lower bitrates.

    \item \textbf{Coding Rate (CR):} The ratio of forward error correction
    bits. Higher CR (e.g., 4/8) increases redundancy and reliability in
    high-attenuation environments.
\end{itemize}

These presets allow the system to adapt to different mission requirements,
ranging from high-speed local telemetry (\textit{Short Turbo}) to
maximum-range emergency messaging (\textit{Long Slow}), as detailed in
Table~\ref{tab:meshtastic_presets}.

\begin{table}[htbp]
\centering
\caption{Meshtastic Modem Presets and LoRa Parameters}
\label{tab:meshtastic_presets}
\small
\begin{tabular}{lcccc}
\toprule
\textbf{Preset Name} & \textbf{BW (\si{\kilo\hertz})} & \textbf{SF} & \textbf{CR} & \textbf{Target} \\
\midrule
Short Turbo   & 500 & 7  & 4/5 & High Speed      \\
Short Fast    & 250 & 7  & 4/5 & Balanced Local  \\
Short Slow    & 250 & 8  & 4/5 & Balanced Fast   \\
Medium Fast   & 250 & 9  & 4/5 & Balanced Mesh   \\
Medium Slow   & 250 & 10 & 4/5 & Robust Mesh     \\
Long Fast*    & 250 & 11 & 4/5 & Default Range   \\
Long Moderate & 125 & 11 & 4/8 & High Range      \\
Long Slow     & 125 & 12 & 4/8 & Maximum Range   \\
\bottomrule
\multicolumn{5}{l}{\scriptsize *Default configuration.}
\end{tabular}
\end{table}

\subsubsection{CSS Modulation and Sub-Noise-Floor Operation.}
The link behaviour is governed by Chirp Spread Spectrum (CSS) modulation. In
urban canyon or emergency rescue scenarios, signals often operate below the
thermal noise floor ($SNR < 0$). LoRa transceivers like the SX1262 can
demodulate signals with negative SNR, down to approximately \SI{-20}{\decibel}
for high spreading factors~\cite{semtech_corporation_sx1276777879_2020}. This
characteristic is fundamental for the experimental design, as the guided-link
attenuator chain must be capable of driving the received signal well below the
noise floor to map the full sensitivity curve.

\subsubsection{SNR vs.\ RSSI: Critical Metric Selection.}
In high-attenuation environments, the Signal-to-Noise Ratio (SNR) becomes a
more critical metric than the Received Signal Strength Indicator (RSSI). While
RSSI measures total power at the receiver front-end, it loses granularity at
the noise floor~\cite{semtech_corporation_an120021_2014}. The actual received
power ($P_{rx}$) is corrected as follows:
\begin{equation}
P_{rx} =
\begin{cases}
RSSI_{packet} + SNR_{packet}, & \text{if } SNR < 0 \\
RSSI_{packet},                & \text{if } SNR \ge 0
\end{cases}
\label{eq:prx-correction}
\end{equation}

This correction is particularly relevant for presets with high SF and narrow BW
(e.g., \textit{Long Slow}), where the receiver operates deep into the
negative-SNR region before link failure occurs.

\subsection{Spectral Verification of LoRa Emissions}

To validate the spectral characteristics of the DUT before the guided-link
experiments, the transmitted signal was captured using two independent
measurement systems. This dual-instrument approach provides both
time-frequency visualisation of the CSS modulation and a calibrated power
spectral density reference.

\begin{figure}[htbp]
  \centering
  \includegraphics[width=0.9\textwidth]{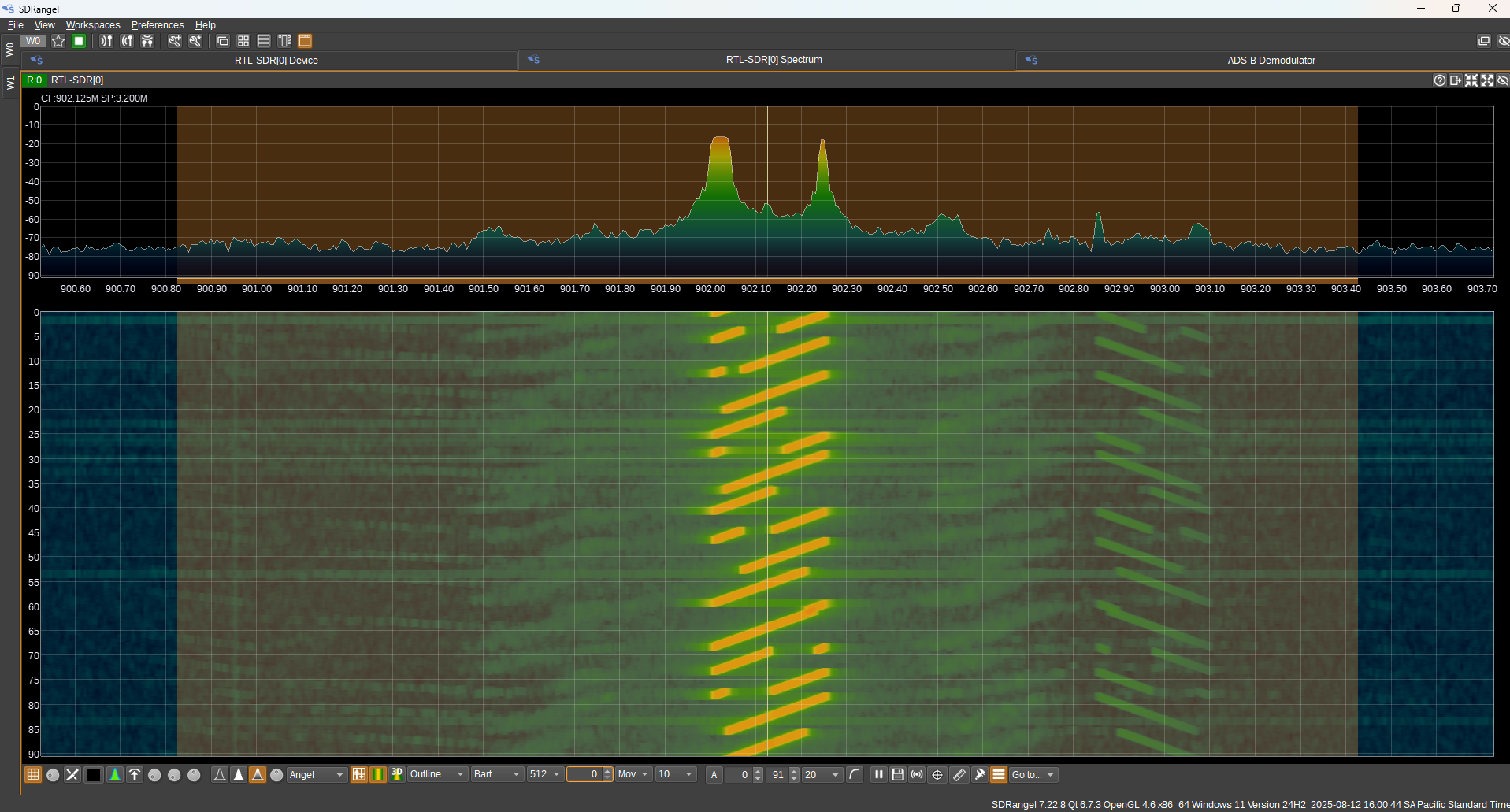}
  \caption{LoRa signal captured with RTL-SDR and SDRangel at
  \SI{902.125}{\mega\hertz} (span: \SI{3.2}{\mega\hertz}). Upper: PSD showing
  transmission peaks. Lower: waterfall display with characteristic CSS chirp
  patterns.}
  \label{fig:spectrum-sdr}
\end{figure}

Figure~\ref{fig:spectrum-sdr} shows the LoRa signal captured using an RTL-SDR
receiver and the SDRangel software, centred at \SI{902.125}{\mega\hertz} with a
span of \SI{3.2}{\mega\hertz}. The upper panel displays the real-time power
spectral density, where two transmission peaks are visible around
\SI{901.9}{\mega\hertz} and \SI{902.2}{\mega\hertz}, reaching approximately
\SI{-20}{\dBm} above a noise floor of \SI{-70}{\dBm} to \SI{-80}{\dBm}. The
lower panel presents the spectrogram (waterfall display), where the
characteristic diagonal chirp patterns of CSS modulation are clearly
identifiable. These frequency sweeps, visible as diagonal bright traces,
confirm that the LoRa transmitter is operating correctly with the expected
modulation scheme.

\begin{figure}[htbp]
  \centering
  \includegraphics[width=0.9\textwidth]{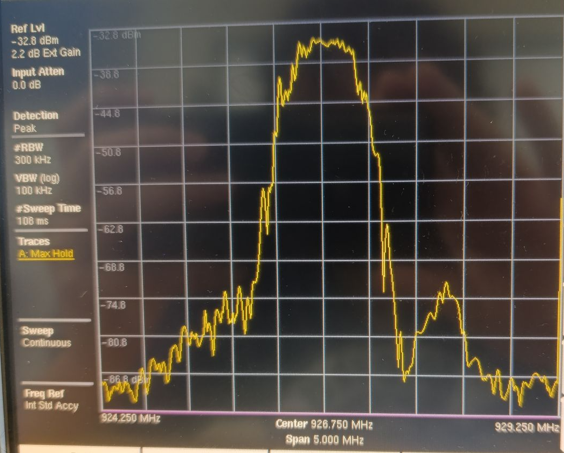}
  \caption{LoRa spectral envelope captured with Anritsu spectrum analyser at
  \SI{926.750}{\mega\hertz} (span: \SI{5}{\mega\hertz}, RBW:
  \SI{300}{\kilo\hertz}, Max Hold). The occupied bandwidth confirms operation
  within the \SI{915}{\mega\hertz} ISM band.}
  \label{fig:spectrum-anritsu}
\end{figure}

Figure~\ref{fig:spectrum-anritsu} presents the spectral measurement obtained
with an Anritsu handheld spectrum analyser, centred at
\SI{926.750}{\mega\hertz} with a \SI{5}{\mega\hertz} span. The instrument was
configured with a resolution bandwidth (RBW) of \SI{300}{\kilo\hertz}, video
bandwidth (VBW) of \SI{100}{\kilo\hertz}, and a sweep time of
\SI{108}{\milli\second} in Max Hold trace mode. The captured spectral envelope
shows the accumulated LoRa signal occupying approximately
\SIrange{925}{928}{\mega\hertz}, with peak power levels near \SI{-35}{\dBm}
(accounting for \SI{2.2}{\decibel} of external gain noted in the instrument
settings). The occupied bandwidth and centre frequency are consistent with the
\SI{915}{\mega\hertz} ISM band channelisation used by Meshtastic in the
Americas region (AU915/US915 frequency plan).

The combination of both measurements---the SDR waterfall confirming CSS chirp
structure and the calibrated analyser verifying power levels and occupied
bandwidth---provides robust evidence that the DUT emits signals conforming to
the LoRa specification prior to entering the guided-link attenuator chain.

\subsection{Experimental Design}

\subsubsection{Variable Classification.}
Table~\ref{tab:variables} explicitly categorises the experimental variables to
ensure reproducibility.

\begin{table}[htbp]
\centering
\caption{Experimental Variable Classification}
\label{tab:variables}
\small
\begin{tabular}{lp{0.3\textwidth}p{0.35\textwidth}}
\toprule
\textbf{Type} & \textbf{Variable} & \textbf{Values / Range} \\
\midrule
\multirow{3}{*}{Independent}
  & Modem Preset    & 8 presets (Table~\ref{tab:meshtastic_presets}) \\
  & TX Power Level  & Low, Medium, Max \\
  & $At_{var}$      & 0--\SI{110}{\decibel} (\SI{1}{\decibel} steps) \\
\midrule
\multirow{3}{*}{Dependent}
  & RSSI & \si{\dBm} \\
  & SNR  & \si{\decibel} \\
  & PER  & \% (per 50-packet burst) \\
\midrule
\multirow{3}{*}{Controlled}
  & Fixed attenuation & $\sum At_{1\text{--}4}$ (constant) \\
  & Temperature        & Ambient (\SIrange{20}{25}{\celsius}) \\
  & Packet payload     & Fixed tag per step \\
\bottomrule
\end{tabular}
\end{table}

To characterise performance across the Meshtastic operational spectrum, a
comprehensive matrix of 42 datasets was executed. This consists of all 8 modem
presets evaluated at three power levels, as summarised in
Table~\ref{tab:test_matrix}. Presets with shorter ToA (\textit{Short Turbo}
through \textit{Long Fast}) were measured in duplicate runs per power level to
assess repeatability, yielding 36 datasets. Presets with longer ToA
(\textit{Long Moderate} and \textit{Long Slow}) were measured once per power
level due to extended acquisition times, contributing 6 additional datasets.

\begin{table}[htbp]
\centering
\caption{Experimental Test Matrix (42 Datasets Total)}
\label{tab:test_matrix}
\small
\begin{tabular}{lccc}
\toprule
\textbf{Meshtastic Preset} & \textbf{Power Levels} & \textbf{Runs/Level} & \textbf{Datasets} \\
\midrule
Short Turbo   & Low, Med, Max & 2 &  6 \\
Short Fast    & Low, Med, Max & 2 &  6 \\
Short Slow    & Low, Med, Max & 2 &  6 \\
Medium Fast   & Low, Med, Max & 2 &  6 \\
Medium Slow   & Low, Med, Max & 2 &  6 \\
Long Fast     & Low, Med, Max & 2 &  6 \\
Long Moderate & Low, Med, Max & 1 &  3 \\
Long Slow     & Low, Med, Max & 1 &  3 \\
\midrule
\multicolumn{3}{r}{\textbf{Total}} & \textbf{42} \\
\bottomrule
\end{tabular}
\end{table}

The variable attenuation ($At_{var}$) was incremented manually using a rotary
step attenuator (0--\SI{110}{\decibel} in \SI{1}{\decibel} steps). For each
attenuation step, a burst of 50 packets was transmitted. After each burst, a
stabilisation period of \SI{5}{\second} was observed before incrementing the
attenuator to the next step. Data was captured through the CP2102 serial
interface at \SI{115200}{baud} and logged to CSV files, recording timestamp,
RSSI, SNR, and a ``payload'' tag correlated with each attenuation step.

The Packet Error Rate (PER) was calculated for each attenuation step as:
\begin{equation}
    PER = \frac{N_{lost}}{N_{sent}} \times 100\%
    \label{eq:per}
\end{equation}
where $N_{sent} = 50$ packets per step and $N_{lost}$ is the number of packets
not received or received with CRC errors. The \textit{sensitivity threshold}
for each modem preset--power combination was defined as the minimum received
power $P_r$ at which $PER \leq 10\%$. Beyond this threshold, the link is
considered unreliable for operational Meshtastic deployment.

\section{Results and Analysis}
\label{sec:results}

This section presents the experimental results obtained from the 42-dataset
test matrix. The analysis focuses on the joint behaviour of RSSI and SNR as a
function of guided-link attenuation across all eight Meshtastic modem presets
at maximum transmission power (effective $+$\SI{21}{\dBm}).

Figures~\ref{fig:performance-metrics} present the complete
measurement results:

\begin{sidewaysfigure}
  \centering
  \includegraphics[width=0.85\textheight]{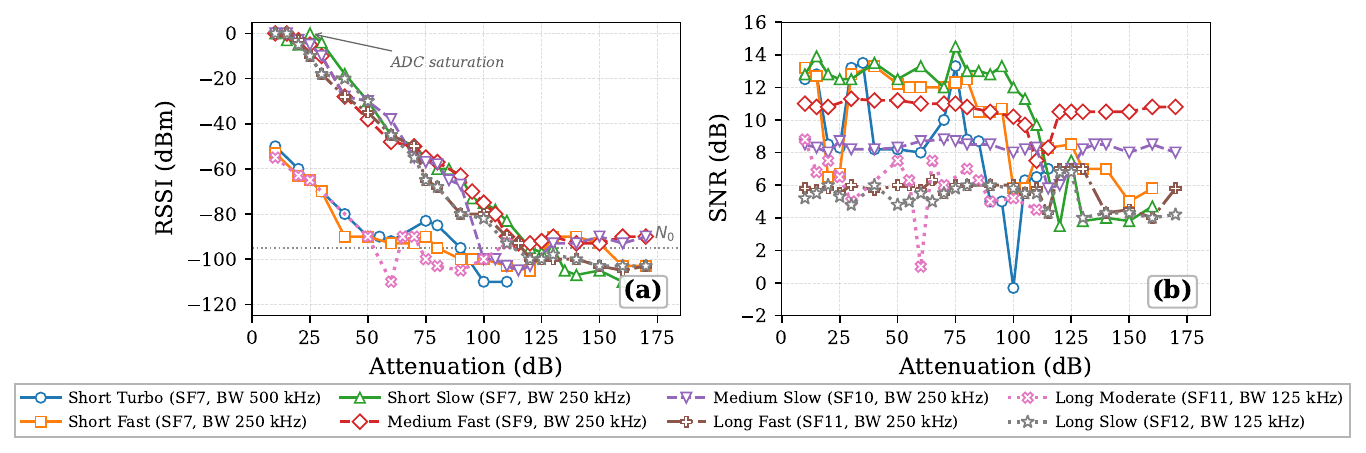}
  
  \vspace{10pt} 
  
  \captionsetup{format=plain, justification=justified, singlelinecheck=false}
  
  \caption{Comparison of Meshtastic modem presets: 
    \textbf{(a)} Received Signal Strength Indicator (RSSI) vs.\ Path Attenuation and 
    \textbf{(b)} Signal-to-Noise Ratio (SNR) vs.\ Path Attenuation. 
    The dashed line $N_0$ indicates the receiver thermal noise floor.}
  \label{fig:performance-metrics}
\end{sidewaysfigure}

\subsection{Spreading Factor and Bandwidth Impact on Sensitivity}

The RSSI data in Fig.~\ref{fig:performance-metrics}(a) reveals distinct behavioural
families, separated primarily by their Spreading Factor~(SF).The \textit{Short} family ($SF7$--$SF8$, $BW \geq \SI{250}{\kilo\hertz}$)
exhibits a quasi-linear RSSI decay of approximately \SI{-1}{\decibel} per
\SI{1}{\decibel} of added attenuation, consistent with the expected response
of a calibrated guided link. \textit{Short Turbo} ($SF7$,
$BW = \SI{500}{\kilo\hertz}$), having the lowest processing gain, reaches the
receiver noise floor first, saturating near \SI{-90}{\dBm} at approximately
\SI{75}{\decibel}. \textit{Short Fast} ($SF7$, $BW = \SI{250}{\kilo\hertz}$)
and \textit{Short Slow} ($SF8$, $BW = \SI{250}{\kilo\hertz}$) collapse
progressively between \SIrange{110}{120}{\decibel}.

The \textit{Medium} family ($SF9$--$SF10$, $BW = \SI{250}{\kilo\hertz}$)
extends reliable reception into the \SIrange{135}{150}{\decibel} range, with
the additional processing gain of $SF9$ and $SF10$ providing approximately
\SIrange{20}{30}{\decibel} of additional margin over the \textit{Short}
presets.The high processing gain \textit{Long} family ($SF11$--$SF12$) demonstrates
markedly extended range. Even at a wider bandwidth, \textit{Long Fast} ($SF11$,
$BW = \SI{250}{\kilo\hertz}$) survives up to \SI{155}{\decibel}.
\textit{Long Slow} ($SF12$, $BW = \SI{125}{\kilo\hertz}$) maximises the link
budget, maintaining packet reception up to approximately \SI{180}{\decibel} of
total attenuation. This additional margin is attributable to the
\SI{3}{\decibel} sensitivity improvement from halving the bandwidth combined
with the maximum CSS processing gain of $SF12$.

\subsection{SNR Stratification and Sub-Noise-Floor Demodulation}

The SNR data in Fig.~\ref{fig:performance-metrics}(b) reveals a clear stratification by
spreading factor. At low attenuation, the presets cluster into tiers: a high
tier (\SIrange{12}{14}{\decibel}) occupied by the $SF7$--$SF8$ presets
(\textit{Short Turbo}, \textit{Short Fast}, and \textit{Short Slow}), a mid
tier (\SIrange{8}{11}{\decibel}) corresponding to $SF9$--$SF11$ at
$BW = \SI{250}{\kilo\hertz}$ (\textit{Medium Fast}, \textit{Medium Slow},
\textit{Long Fast}), and a lower initial tier (\SIrange{5}{7}{\decibel}) for
the narrowband $BW = \SI{125}{\kilo\hertz}$ presets (\textit{Long Moderate},
\textit{Long Slow}).

A defining characteristic of the LoRa physical layer is explicitly captured in
this dataset: the ability to demodulate signals deeply embedded in thermal
noise. As attenuation increases beyond \SI{120}{\decibel}, the SNR for
high-SF presets crosses the zero-threshold and descends linearly into the
negative regime. \textit{Long Slow} reaches absolute SNR values of
approximately \SI{-18}{\decibel} before link failure, demonstrating that the
CSS processing gain effectively preserves demodulation capability near the
SX1262's theoretical limit (\SI{-20}{\decibel} for $SF12$).The data also suggests that link failure for high-SF presets is \textit{abrupt}
rather than gradual. The transition from reliable reception to complete link
loss occurs while the system still maintains a tight \SIrange{2}{4}{\decibel}
margin above the absolute theoretical demodulation limit, confirming that fade
margins in the order of \SI{5}{\decibel} are required for stable network
planning.

\subsection{Measurement Artefacts}

Two systematic artefacts were identified in the dataset, providing important
bounds for signal interpretation.

\subsubsection{Front-End ADC Saturation.}
Several presets report $RSSI \approx \SI{0}{\dBm}$ at low attenuation levels
(\SIrange{0}{40}{\decibel}), visible in the upper-left region of
Fig.~\ref{fig:performance-metrics}(a). This indicates ADC saturation: the received signal
exceeds the dynamic range of the SX1262 receiver, causing the RSSI register to
clip. These data points do not represent actual propagation loss and must be
excluded from path-loss regression models.

\subsubsection{RSSI Plateau at the Noise Floor.}
Beyond the sensitivity limit, all surviving presets exhibit an RSSI plateau
near \SI{-100}{\dBm} that persists across extreme attenuation levels. In this
regime, the RSSI register reflects the receiver's thermal noise power rather
than the signal energy. Consequently, the corrected received power must be
calculated using the SNR-adjusted formula~(Eq.~\ref{eq:prx-correction}),
validating the theoretical prediction that SNR is the sole reliable link
quality metric in the sub-noise-floor operating regime.

\subsection{Anomalous Behaviour of Long Moderate}
\label{subsec:anomalies}

The \textit{Long Moderate} preset ($SF11$, $BW = \SI{125}{\kilo\hertz}$,
$CR = 4/8$) exhibits a highly anomalous discontinuity. As seen in
Figs.~\ref{fig:performance-metrics}(a) and~\ref{fig:performance-metrics}(b), the link fails
abruptly near \SI{80}{\decibel} of attenuation. Crucially, at the point of
failure, the preset still maintains an extraordinarily robust SNR of
approximately $+\SI{10}{\decibel}$. Because this abrupt drop does not follow
the physical layer progression seen in \textit{Long Slow} ($SF12$, same BW)
nor \textit{Long Fast} ($SF11$, wider BW), it strongly suggests a specific
interaction between $SF11$, $CR = 4/8$, and the firmware's Channel Activity
Detection (CAD) or packet timeout mechanisms, rather than a true RF sensitivity
limitation.

\section{Discussion}
\label{sec:discussion}

The experimental results presented in Section~\ref{sec:results} provide the
empirical foundation for selecting the appropriate Meshtastic configuration
according to the deployment scenario. This section analyses how the choice of
modem preset directly determines the achievable inter-node distance and,
consequently, the node density required to guarantee mesh connectivity over a
given coverage area.

\subsection{Operational Regimes and Configuration Selection}

Three operational regimes can be identified from the sensitivity thresholds:

\begin{itemize}
    \item \textbf{High-density, low-latency applications}
    (\textit{Short Turbo}, \textit{Short Fast}, \textit{Short Slow}): suitable
    for dense urban IoT sensor networks, building automation, or campus-scale
    telemetry where nodes are closely spaced ($<$\SI{500}{\meter}) and frequent
    data updates are required. The high throughput and low ToA minimise channel
    occupancy per transmission, enabling a larger number of nodes to coexist
    within the same mesh without excessive collision probability. The
    $SF7$--$SF8$ presets tolerate \SIrange{110}{120}{\decibel} of path loss,
    which in a typical dense urban environment (path loss exponent
    $n \approx 3.5$--$4$) corresponds to inter-node distances in the order of
    \SIrange{100}{500}{\meter}.

    \item \textbf{Balanced urban mesh} (\textit{Medium Fast}, \textit{Medium
    Slow}, \textit{Long Fast}): appropriate for neighbourhood-scale smart city
    deployments---environmental monitoring, traffic sensing, public safety
    alerts---where moderate inter-node distances (\SIrange{0.5}{2}{\kilo\meter})
    balance coverage and throughput. \textit{Long Fast}, as the default
    Meshtastic configuration, represents the most versatile trade-off for
    general-purpose urban mesh networks, sustaining links up to
    \SI{155}{\decibel} of attenuation. \textit{Medium Fast} is notable for
    maintaining a uniform SNR margin of approximately \SI{10}{\decibel} across
    its operational range, making it particularly robust against temporal fading
    variations.

    \item \textbf{Maximum-range emergency networks} (\textit{Long Slow}):
    designed for disaster response, rural coverage extension, or
    search-and-rescue operations where inter-node distances may exceed
    \SIrange{2}{10}{\kilo\meter}. The reduced throughput and high ToA are
    acceptable trade-offs when the primary objective is to establish any
    communication link over the largest possible area with the minimum number
    of available nodes. The \SI{180}{\decibel} link budget of \textit{Long
    Slow} enables theoretical LOS ranges exceeding \SI{10}{\kilo\meter} even
    with conservative fade margins.
\end{itemize}

It should be noted that the anomalous behaviour of \textit{Long Moderate}
(Section~\ref{subsec:anomalies}) currently limits its practical applicability
despite its favourable theoretical parameters. Until the root
cause---likely firmware-related---is resolved, \textit{Long Slow} remains the
recommended preset for maximum-range deployments.

\subsection{Node Density and Economic Implications}

The \SIrange{60}{70}{\decibel} span between the \textit{Short Turbo} and
\textit{Long Slow} sensitivity thresholds translates directly into node density
requirements for a given coverage area. Table~\ref{tab:density-implications}
illustrates the estimated inter-node distances for a representative dense urban
scenario.

\begin{table}[htbp]
\centering
\caption{Estimated Inter-Node Distance and Density for Dense Urban Deployment
($n = 3.5$, $P_{Tx} = +\SI{21}{\dBm}$)}
\label{tab:density-implications}
\small
\begin{tabular}{lccc}
\toprule
\textbf{Regime}
  & \textbf{Max Att.\ (dB)}
  & \textbf{Est.\ Range (m)}
  & \textbf{Nodes/km\textsuperscript{2} (approx.)} \\
\midrule
Short (SF7--8)   & 110--120 & 200--400   & 25--80 \\
Medium (SF9--10) & 135--150 & 500--1500  & 3--12  \\
Long (SF11--12)  & 155--180 & 1500--5000 & 1--3   \\
\bottomrule
\multicolumn{4}{l}{\scriptsize Ranges assume $f = \SI{915}{\mega\hertz}$,
free-space + urban excess loss.} \\
\multicolumn{4}{l}{\scriptsize Actual values depend on building density and
terrain.}
\end{tabular}
\end{table}

A \textit{Short Turbo} deployment covering \SI{1}{\kilo\meter\squared} may
require 25 to 80 nodes, while the same area under \textit{Long Slow} could be
served by as few as one to three strategically placed nodes---at the cost of
message latency increasing from seconds to minutes. For a municipal risk
management agency planning a city-wide early warning system, this represents a
trade-off between hardware investment (more nodes, lower unit cost per message)
and operational simplicity (fewer nodes, higher per-node criticality).

The guided-link sensitivity thresholds established in Section~\ref{sec:results}
provide the baseline link budget from which planners can subtract the expected
path loss---using models such as ITU-R P.1411 for urban microcells or the
Okumura-Hata model for wider areas---to estimate the maximum inter-node spacing
for each preset in a specific urban morphology. For Latin American cities,
where building construction (reinforced concrete, brick masonry) introduces
additional NLoS penetration losses at \SI{915}{\mega\hertz}, a conservative
margin of \SIrange{10}{15}{\decibel} beyond the measured thresholds is
recommended.

\section{Conclusions}
\label{sec:conclusions}

This work presented a guided-link experimental methodology for the
deterministic characterisation of Meshtastic nodes operating in the
\SI{915}{\mega\hertz} ISM band. By replacing the wireless channel with a
calibrated attenuator cascade, the study isolated the LoRa physical layer from
stochastic propagation effects, enabling repeatable sensitivity and PER
measurements across the full operational envelope of the SX1262 transceiver.

The evaluation of all eight modem presets at three power levels (42 datasets)
revealed three distinct performance tiers governed primarily by Spreading
Factor: the \textit{Short} family ($SF7$--$SF8$) failed between
\SIrange{110}{120}{\decibel}, the \textit{Medium} family ($SF9$--$SF10$)
sustained links up to \SIrange{135}{150}{\decibel}, and the \textit{Long}
family ($SF11$--$SF12$) reached \SI{180}{\decibel}---a \SIrange{60}{70}{\decibel}
advantage that translates to multiple orders of magnitude in propagation
distance. The SNR analysis confirmed sub-noise-floor demodulation down to
\SI{-18}{\decibel} for $SF12$, with abrupt link failure within
\SIrange{2}{4}{\decibel} of the theoretical limit, establishing that fade
margins of \SI{5}{\decibel} suffice for stable network planning.

These thresholds enabled the identification of three operational
regimes---high-density IoT, balanced urban mesh, and maximum-range
emergency---each with distinct node density and latency characteristics
(Section~\ref{sec:discussion}). The analysis provides network designers and
municipal risk management agencies with quantitative criteria to dimension
Meshtastic deployments according to coverage requirements, budget constraints,
and urban morphology.

Two measurement artefacts were documented (ADC saturation and noise floor
plateau), and an anomalous firmware-level failure of the \textit{Long Moderate}
preset was identified, warranting further investigation.

Meshtastic represents a vital convergence between the sub-noise-floor
capabilities of SX1262 transceivers and the need for communication resilience
in smart cities. For Latin American urban contexts---where licence-free
\SI{915}{\mega\hertz} operation, low hardware cost, and autonomous power supply
enable rapid deployment---the configuration-dependent design guidelines
presented here constitute a practical planning framework for off-grid emergency
and IoT networks.

\subsection{Future Work}

The results open several research directions:

\begin{enumerate}
    \item \textbf{Outdoor propagation validation:} Field measurements in
    representative urban environments (dense urban, suburban, urban canyon),
    with particular emphasis on NLoS penetration losses through Latin American
    building materials at \SI{915}{\mega\hertz}.

    \item \textbf{Flooding protocol scalability:} Evaluation of Meshtastic's
    managed flooding in dense NLoS scenarios, and exploration of hybrid routing
    strategies (selective forwarding, hierarchical clustering) for formal
    adoption by risk management entities.

    \item \textbf{Node density optimisation tool:} Development of a planning
    tool mapping empirical sensitivity thresholds to optimal node placement for
    a given urban morphology, terrain, and QoS requirements.

    \item \textbf{IoT platform integration:} Benchmarking of dual-mode
    operation (autonomous mesh / cloud-connected) with urban management
    platforms under realistic multi-hop traffic loads.

    \item \textbf{Dynamic channel emulation:} Extension of the guided-link
    methodology to include multipath fading profiles and Doppler spread via
    hardware channel emulator.
\end{enumerate}

\bibliographystyle{splncs04}
\bibliography{meshtastic_1}

\end{document}